\documentclass{article}
\usepackage[preprint]{spconf}
\usepackage{amsmath, graphicx}
\usepackage[hidelinks]{hyperref}

\usepackage{cleveref}

\usepackage{booktabs}

\usepackage{multirow}

\usepackage{xcolor}

\usepackage[acronym]{glossaries} %nonumberlist           

\newacronym{ram}{RAM}{Random Access Memory}
\newacronym{nn}{NN}{Neural Network}
\newacronym{ann}{ANN}{Artificial Neural Network}
\newacronym{dnn}{DNN}{Deep Neural Network}
\newacronym{mlp}{MLP}{Multilayer Perceptron}
\newacronym{cnn}{CNN}{Convolutional Neural Network}
\newacronym{bnn}{BNN}{Bayesian Neural Network}
\newacronym{rnn}{RNN}{Recurrent Neural Network}
\newacronym{crnn}{CRNN}{Convolutional Recurrent Neural Network}
\newacronym{moco}{MoCo}{Momentum Contrast}
\newacronym{kws}{KWS}{Keyword Spotting}
\newacronym{asr}{ASR}{Automatic Speech Recognition}
\newacronym{hmm}{HMM}{Hidden Markov Model}
\newacronym{kwt}{KWT}{Keyword Transformer}
\newacronym{ema}{EMA}{Exponential Moving Average}
\newacronym{lvcsr}{LVCSR}{Large-vocabulary Continous Speech Recognition}
\newacronym{gmm}{GMM}{Gaussian Mixture Model}
\newacronym{mfcc}{MFCC}{Mel-frequency cepstral coefficient}
\newacronym{dct}{DCT}{Discrete Cosine Transform}
\newacronym{lstm}{LSTM}{Long Short-term Memory}
\newacronym{gru}{GRU}{Gated Recurrent Unit}
\newacronym{wandb}{WandB}{Weights and Biases}
\newacronym{hpc}{HPC}{High Performance Computing}
\newacronym{gelu}{GELU}{Gaussian Error Linear Unit}
\newacronym{vicreg}{VICReg}{Variation-Invariance-Covariance regularization}
\newacronym{tcl}{TCL}{Time Contrastive Learning}
\newacronym{byol}{BYOL}{Bootstrap Your Own Latent}
\newacronym{apc}{APC}{Autoregressive Predictive Coding}

\usepackage{siunitx}						
\DeclareSIUnit\byte{B}

\title{Improving Label-deficient Keyword Spotting Through Self-supervised Pretraining}

\copyrightnotice{\copyright\ IEEE 2023}
\toappear{To appear in {\it Proc.\ ICASSP2023 SASB Workshop, June 10th, 2023, Rhodes, Greece}}

\name{Holger Severin Bovbjerg$^{\star}$, Zheng-Hua Tan$^{\star \dagger}$} % \thanks{Thanks to XYZ agency for funding.}
\address{$^{\star}$Department of Electronic Systems, Aalborg University, Denmark\\
$^{\dagger}$Pioneer Centre for AI, Denmark\\
\texttt{\{hsbo,zt\}@es.aau.dk}}
\begin{document}
%\ninept
%
\maketitle
\begin{abstract}
Keyword Spotting (KWS) models are becoming increasingly integrated into various systems, e.g. voice assistants. 
To achieve satisfactory performance, these models typically rely on a large amount of labelled data, limiting their applications only to situations where such data is available. 
Self-supervised Learning (SSL) methods can mitigate such a reliance by leveraging readily-available unlabelled data.
Most SSL methods for speech have primarily been studied for large models, whereas this is not ideal, as compact KWS models are generally required.
This paper explores the effectiveness of SSL on small models for KWS and establishes that SSL can enhance the performance of small KWS models when labelled data is scarce.
We pretrain three compact transformer-based KWS models using Data2Vec, and fine-tune them on a label-deficient setup of the Google Speech Commands data set.
It is found that Data2Vec pretraining leads to a significant increase in accuracy, with label-deficient scenarios showing an improvement of \SIrange{8.22}{11.18}{\percent} absolute accuracy. 
\end{abstract}
\section{Introduction}
\label{sec:intro}
Common for personal assistants like Google Assistant and Apple's Siri is that they make use of an \gls*{asr} system, which is activated by a smaller \gls*{kws} system in order to save resources when the \gls*{asr} system is not needed \cite{ivan_deep_spoken_keyword}. 
Modern deep learning based \gls*{kws} models have improved the accuracy of \gls*{kws} systems. However, they need to be trained on a large amount of labelled data to generalize well and obtaining properly labelled speech data is a labour-intensive and costly process, especially for low-resource languages. 

Recently, self-supervised learning methods have shown to be able to learn strong representations from unlabelled data, yielding good performance on a number of downstream tasks, including \gls*{kws}, when fine-tuned on a limited amount of labelled data.
However, current studies mainly focus on developing universal speech models \cite{baevski_wav2vec2, Chen_WavLMLS}, which are trained on large speech corpuses such as Librispeech \cite{librispeech} or LibriLight \cite{librilight}, with the goal of obtaining a model that can perform well for multiple downstream tasks.
These large models are commonly evaluated on benchmarks like SUPERB \cite{Yang_SUPERBSP}, requiring fine-tuning on multiple downstream tasks. 
Consequently, training these models require numerous high-end GPUs and often several weeks of training, making training these models infeasible in many cases, e.g., due to limited time or restricted computing resources.
Additionally, for many use cases, such as \gls*{kws} for voice assistants, it is desirable that the models are small and efficient \cite{ivan_deep_spoken_keyword}.

While knowledge distillation \cite{hinton_distilling} has been investigated for transferring the representations learned by a large model to a smaller model \cite{Gu_Liu, Fang_SEED, chen_kornblith}, such methods do not deal with the problem of the necessity of training a large model initially.
One study used a contrastive type of SSL method to train smaller models without distillation from a large pretrained model and found that, contrary to former assumptions, small models are able to solve the self-supervised pretext tasks without overfitting \cite{efficacy_of_small_self_supervised_contrastive_models}. 
Additionally, they were able to improve the performance of five different small image recognition models, ranging from 2.5 to 11 million parameters, suggesting that training small self-supervised models is feasible.
Other work found that the learned parameters of large speech models suffer from redundancy across layers, and proposed the use of weight sharing to reduce parameter redundancy and the network size \cite{Chi_AudioAA}. 

In this paper, we investigate the adaption of the general \emph{non-contrastive SSL} framework Data2Vec \cite{data2vec} to improve \gls*{kws} performance in label-deficient scenarios. 
We implement three variations of the \gls*{kwt} model \cite{berg_KWT}, varying from 600k to 5.4M parameters, and pretrain the models using Data2Vec. 
The models are evaluated on a label-deficient setup of the Google Speech Commands data set \cite{speechcommandsv2} with only \SI{20}{\percent} labelled data for supervised training, and the results show the following: 
\begin{enumerate}
    \setlength\itemsep{0.2em}
    \item Self-supervised pretraining significantly improves the \gls*{kws} performance for all three models when the amount of labelled data is limited, indicating that self-supervised learning can also be beneficial for small models.
    \item All three pretrained and fine-tuned models achieve similar performance to models trained on \SI{100}{\percent} labelled data, while using \SI{80}{\percent} data as unlabelled data for self-supervised pretraining, and only \SI{20}{\percent} labelled data for fine-tuning.
    \item A significant performance improvement from self-supervised pretraining is also seen when using a larger out-of-domain data set for pretraining.
    \item Fine-tuning the entire pretrained model is necessary.
\end{enumerate}

The source code used to produce the results of this paper is made publicly available\footnote{\url{https://github.com/HolgerBovbjerg/data2vec-KWS}}.

\section{Methodology and Data Sets}

\subsection{Keyword spotting model}
\begin{figure}[htb]
    \centering
    \includegraphics{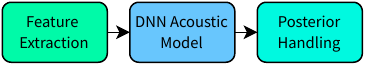}
    \caption{A typical deep KWS system.}
    \label{fig:typical_kws_system}
\end{figure}

A deep \gls*{kws} system can typically be divided into three parts, namely a feature extractor, a \gls*{dnn} acoustic model and posterior handling as shown in \Cref{fig:typical_kws_system}. 

As the Data2Vec framework uses a transformer encoder, we choose to use the transformer based \gls*{kwt} model \cite{berg_KWT}, which has achieved state-of-the-art performance on the Google Speech Commands \gls*{kws} benchmark.
The \gls*{kwt} model is based on a vision transformer \cite{ViT_paper}, substituting images with \glspl*{mfcc}. 

\Glspl*{mfcc} are extracted using a window length of 480, a hop length of 160, and we use 40-dimension \gls*{mfcc} features.
Each \gls*{mfcc} vector is then passed through a linear layer to yield embeddings matching the transformer input dimension.

\begin{figure}[htb]
    \centering
    \includegraphics{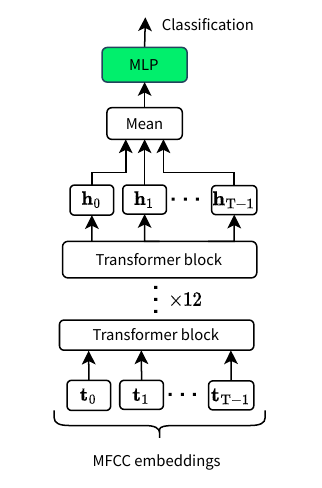}
    \caption{The keyword spotting model.}
    \label{fig:KWT_model}
\end{figure}

The acoustic model consists of 12 transformer blocks, followed by a \gls*{mlp} classification head. 
The \gls*{kwt} model \cite{berg_KWT} concatenates a $\mathrm{CLS}$ (i.e., classification) token to the input, yielding a global encoding of all time steps used as input for the classification head.  
However, we found that using the mean of the encodings of each time step (i.e., MFCC vector) as the input for the classification head yields better performance, both with and without self-supervised pretraining. 
The classification head simply outputs posteriors over the $N_\mathrm{c}$ keyword classes. 
An illustration of the \gls*{kws} model used in this study is seen in \Cref{fig:KWT_model}.

Following \cite{berg_KWT}, the encoder dimension of each transformer block, $d$, is set such that $\frac{d}{k}=64$, where $k$ is the number of attention heads in the multi-head attention block. 

\subsection{Data2Vec pretraining}
\begin{figure}[htb]
    \centering
    \includegraphics{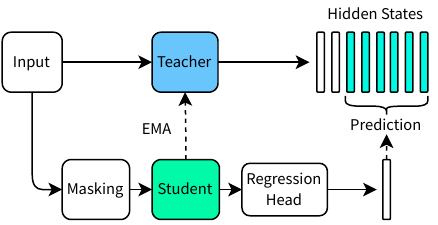}
    \caption{The Data2Vec framework.}
    \label{fig:data2vec_framework}
\end{figure}

Data2Vec \cite{data2vec} uses a student and a teacher model which are identical transformer encoders and the teacher model weights are an \gls*{ema} of the student model weights.
A general overview of the Data2Vec framework is presented in \Cref{fig:data2vec_framework}. 
As illustrated, the teacher model encodes the full input, while the student model encodes a masked version of the input.

The learning objective in the Data2Vec framework is to minimize the difference between the student prediction $f_t(x)$ and the target $y_t$, which is formed from the $K$ last hidden representations of the teacher model as %seen in \eqref{eq:data2vec_target}.

\begin{equation}
    y_t = \frac{1}{K} \sum_{l=L-K+1}^L \hat{\mathbf{h}}_t^l \label{eq:data2vec_target}
\end{equation}
where $y_t$ is the target at time step $t$, $L$ is the total number of transformer blocks, and $\hat{\mathbf{h}}_t^l$ is the normalized hidden state representation from transformer block $l$ at time step $t$.

% Target normalization serves to prevent the model from model collapse, i.e., finding a trivial solution such as a constant representation \cite{data2vec, understanding_model_collapse}. 

\subsection{Data set}
For training and evaluation of the \gls*{kws} model, we use a label-deficient version of the Google Speech Commands V2 data set \cite{speechcommandsv2}, consisting of 105829 \SI{1}{\second} recordings of 35 different keywords.
To simulate a label-deficient scenario where most of the data is unlabelled and only a small amount of labelled data is available, we randomly split the original training set such that \SI{80}{\percent} (\SI{18.8}{\hour}) of the original training set is set aside for unlabelled pretraining and the remaining \SI{20}{\percent} (\SI{4.7}{\hour}) is used as the labelled training set. The resulting splits and their number of examples are summarized in \Cref{tab:speech_commands_splits}. 

\begin{table}[htb]
\centering
\caption{Label-deficient Google Speech Commands V2 splits.}
\label{tab:speech_commands_splits}
\begin{tabular}{@{}lllll@{}}
\toprule
Split    & Pretrain    & Train       & Validation  & Test        \\ \midrule
Examples & \num{67874} & \num{16969} & \num{9981} & \num{11005} \\ \bottomrule
\end{tabular}
\end{table}

In addition to the Speech Commands pretraining set, we also carry out experiments using Librispeech \cite{librispeech} \SI{100}{\hour} clean training set for unlabelled pretraining, in order to test the effect of using a larger data set that is not domain-specific for pretraining.

\section{Experiments}
To gain insight into how model size influences the performance with and without pretraining, we use three models of varying sizes. Following \cite{berg_KWT}, the number of attention heads in the transformer blocks, $k$, is varied from 1 to 3, and the encoder dimension, $d$, from 64 to 192, yielding three models, KWT-1, KWT-2 and KWT-3, with \num{0.6d6}, \num{2.4d6} and \num{5.4d6} parameters, respectively.

The models were implemented in PyTorch and experiments have been carried out on a virtual machine with 10 CPUs, \SI{40}{\giga\byte} RAM and one NVIDIA T40 GPU with \SI{16}{\giga\byte} \gls*{ram}. 
With this setup, pretraining and fine-tuning of the largest model takes approximately \SI{10}{\hour} and \SI{1}{\hour}, respectively.

\subsection{Baseline}
In order to evaluate the benefit of using Data2Vec pretraining, a baseline without pretraining has been established. 
All three baseline models are trained with supervision on the label-deficient Speech Command training set containing \SI{20}{\percent} of the original training set data.

During the supervised training of the baseline model, we apply SpecAugment \cite{specaugment} randomly masking blocks in both time and feature dimension.
We train the \gls*{kwt} model for 140 epochs using a batch size of 512, and use cross entropy as the learning objective. 
Weights are updated using the AdamW \cite{AdamW_optimizer} optimizer with a learning rate of \num{1d-3} and a weight decay of 0.1. We use a learning rate schedule with 10 epochs linear warmup followed by cosine annealing.  
The models are trained and evaluated on the 35 keyword classification setting of the Speech Commands V2 data set.

A simple classification accuracy metric is used as the performance metric for evaluation of the \gls*{kws} system, as the Speech Commands V2 data set is rather balanced in terms of different keywords \cite{speechcommandsv2, ivan_deep_spoken_keyword}.  

\subsection{Pretraining and fine-tuning}
During pretraining, the classification head in \Cref{fig:KWT_model} is replaced by a linear regression head, which predicts the hidden state representations of the teacher model.  
Following \cite{data2vec}, we use a time-domain masking strategy also used in Wav2Vec \cite{wav2vec}.
Specifically, \gls*{mfcc} vectors are sampled with a probability $p_\mathrm{mask}$, and the following $N_\mathrm{mask}=10$ \gls*{mfcc} vectors are replaced by a $\mathrm{MASK}$ token embedding such that approximately \SI{65}{\percent} of the time steps are masked. 
During pretraining, a mask is generated for each input sample, and the student model then encodes the masked embeddings, while the teacher model encodes the unmasked embeddings. 

Most of the hyperparameters for pretraining are chosen according to the original Data2Vec study \cite{data2vec}, with some slight alterations due to differences in data and hardware setup. 
Specifically, we set the batch size to 512, the number of EMA decay annealing steps to 1000 and training for 200 epochs.
We update the student weights using a mean squared error (MSE) loss and Adam \cite{Kingma_Adam} optimizer with a learning rate of \num{0.5d-3} and a weight decay of 0.01, using a 1-cycle learning rate schedule \cite{Smith_1cycle}.

After Data2Vec pretraining, the regression head is replaced with the original classification head to fine-tune the model for \gls*{kws}. All three models are fine-tuned on the label-deficient Speech Command training set containing \SI{20}{\percent} of the original training set data. 
During fine-tuning, we load a pretrained model and fine-tune the whole model using the same procedure as used for the baseline. 

\section{Results}
\Cref{tab:kwt_results} shows the test set accuracies for both the baseline models and the fine-tuned models as a comparison. %The test set accuracies for the baseline models are seen in \Cref{tab:kwt_results}, which, as we will show later, are significantly worse than the models trained using \SI{100}{\percent} of labelled data from the original Speech Commands training set.
All models in this table have been trained or fine-tuned for \gls*{kws} using the label-deficient training set of \Cref{tab:speech_commands_splits}.

\begin{table}[t]
\centering
\caption{Summary of results for the three KWT models of different settings. All experiments used \SI{20}{\percent} labelled data (i.e. the label-deficient setting). Baseline denotes models only trained on the label-deficient Speech Commands training set without pretraining. SC denotes Data2Vec pretraining using Speech Commands pretraining set, and LS denotes pretraining using Librispeech 100-hour clean training set.} 
\vspace{0.1cm}
\label{tab:kwt_results}
\begin{tabular}{@{}lllll@{}}
\toprule
      &        &  & \multicolumn{2}{c}{Data2Vec pretraining} \\ 
\cmidrule{4-5}
Model & Parameters & Baseline & SC                  & LS                 \\
\midrule
KWT-1 & 600k & 0.8572   & 0.9394              & \textbf{0.9426}    \\
KWT-2 & 2.4M & 0.8584   & \textbf{0.9507}     & 0.9455             \\
KWT-3 & 5.4M & 0.8411   & \textbf{0.9527}     & 0.9476             \\ \bottomrule
\end{tabular}
\end{table}

The models pretrained using Speech Commands pretraining data show absolute improvements in accuracy over the baseline between \SIrange{8.22}{11.18}{\percent}, with the KWT-3 model having the best accuracy, achieving a score of \SI{95.29}{\percent}. 
The two smaller models, KWT-1 and KWT-2, also achieve a similar performance improvement, with an accuracy of \SI{93.94}{\percent} and \SI{95.07}{\percent} respectively.

As shown in \Cref{tab:kwt_results}, using \SI{100}{\hour} clean Librispeech data for pretraining yields very similar performance to that of the models pretrained using Speech Commands data (\SI{18.8}{\hour}), with minor differences in accuracy between them in the range from \SI{0.32}{\percent} to \SI{0.52}{\percent}. 
This suggests that despite using relatively small KWS models, the learned models from pretraining on Librispeech data generalize well to the Speech Commands data set. 

Generally, the results demonstrate that for all three model sizes, Data2Vec pretraining can significantly improve \gls*{kws} performance in the label-deficient scenarios, even when pretrained on a relatively small amount of unlabelled data.
Additionally, pretraining on the \SI{100}{\hour} Librispeech data set yields a similar improvement, showing that the pretraining data does not need to be domain-specific. 
Overall, these results indicate that self-supervised pretraining can also be beneficial for small \gls*{kws} models in label-deficient scenarios, while also not relying on gathering domain-specific data for pretraining.

\subsection{Ablation studies}
Following the initial experiments, we now investigate the use of the pretrained model as a feature extractor, as well as scaling up the amount of labelled data to the full original training set.
These results are presented in \Cref{tab:kwt_results_ablation}.

We observe that fixing the pretrained model weights, using the fine-tuned model as a feature extractor, and only training the classification head, yields drastically worse performance, while all three models show a significant improvement in performance, when fine-tuning all weights. 
This suggests that the learned representations during pretraining are not strong enough to use directly as input features for the linear classifier, thus allowing all weights to be updated is necessary to achieve good performance. 

In addition, we also train the models on \SI{100}{\percent} of labelled data from the original training set, to use as a performance reference. Interestingly, we find that training the models on \SI{100}{\percent} labelled data does not outperform the Data2Vec pretrained models fine-tuned on \SI{20}{\percent} labelled data only, with the latter  showing similar or even slightly better performance. All baseline models presented in \Cref{tab:kwt_results} perform significantly worse than the models trained on the full training set, which is expected, due to the limited amount of labelled training data (\SI{20}{\percent}) for the baseline models.

We also find that when fine-tuning the models pretrained on \SI{100}{\hour} Librispeech data using the full (i.e. {\SI{100}{\percent}} of) Speech Commands training set, a further improvement in \gls*{kws} performance in seen. 
Here, all models achieve an accuracy above \SI{97}{\percent}, with the larger KWT-2 and KWT-3 models achieving the largest increases in accuracy relative to no pretraining. 

\begin{table}[t]
\centering
\caption{Summary of results for feature extraction and for training on the full original training set. "-FE" denotes that the pretrained models were used for feature extraction and their weights were fixed. "Full" indicates models trained or fine-tuned on the full Speech Commands V2 training set (i.e. \SI{100}{\percent} labelled) instead of the label-deficient training set (\SI{20}{\percent} only).}
\vspace{0.1cm}
\label{tab:kwt_results_ablation}
\begin{tabular}{@{}llll@{}}
\toprule
Model & SC-FE    & Full       & Full + LS pretrain            \\
\midrule
KWT-1 & 0.4292   & 0.9638     & 0.9713             \\
KWT-2 & 0.4974   & 0.9498     & 0.9716             \\
KWT-3 & 0.4960   & 0.9079     & 0.9716             \\ \bottomrule
\end{tabular}
\end{table}

\section{Conclusion}
In this paper, we investigated the use of the self-supervised learning method, Data2Vec, in order to improve the performance of \gls*{kws} models in label-deficient scenarios. 
We implemented a self-supervised keyword spotting system using three variations of the \gls*{kwt} model which we pretrained using Data2Vec, and tested the system on a label-deficient setup of the Google Speech Commands data set.

The results show that pretraining using Data2Vec significantly improved the \gls*{kws} performance for all three models, with absolute improvements in test set accuracy between \SIrange{8.22}{11.18}{\percent}. 
Moreover, when fine-tuned using only \SI{20}{\percent} of the original labelled training data, the pretrained models achieved performance comparable to models trained on the full training set, regardless of whether the models were pretrained on domain-specific data.

These significant improvements in accuracy show that self-supervised pretraining is not limited to large general speech models but can also greatly improve the performance of relatively small \gls*{kws} models, when labelled data is limited. 
Further work involves investigating how self-supervised pretraining affects the noise-robustness of \gls*{kws} models, and obtaining deeper insights on the limitations of self-supervised pretraining for small models.

\newpage

\bibliographystyle{IEEEbib}
\bibliography{references}

\end{document}